\documentclass[twocolumn,aps,prb]{revtex4-1}
\usepackage{graphicx}
\usepackage{amsmath}
\usepackage{url}
\usepackage{float}
\usepackage[english]{babel}
\usepackage{chngcntr}
\usepackage{natbib}
\usepackage{textcase}
\usepackage{hyperref}
\usepackage{tikz}
\usepackage{xcolor}
\usepackage[percent]{overpic}
\usepackage{chngcntr}

\hypersetup{
    colorlinks,
    citecolor=blue,
    filecolor=blue,
    linkcolor=blue,
    urlcolor=blue
}

\begin{document}

\title{Electric dipole spin resonance in systems with a valley dependent $g$-factor}

\author{Marko J. Ran\v{c}i\'{c}}
\author{Guido Burkard}
\affiliation{Department of Physics, University of Konstanz, D-78457 Konstanz, Germany}

%\date{\today}

\begin{abstract}
In this theoretical study we qualitatively and quantitatively investigate the electric dipole spin resonance (EDSR) in a single Si/SiGe quantum dot in the presence of a magnetic field gradient, e.g., produced by a ferromagnet.
We model a situation in which the control of electron spin states is achieved by applying an oscillatory electric field, inducing real-space oscillations of the electron inside the quantum dot.
One of the goals of our study is to present a microscopic theory of valley dependent $g$-factors in Si/SiGe quantum dots and investigate how valley relaxation combined with a valley dependent $g$-factor leads to a novel electron spin dephasing mechanism. 
Furthermore, we discuss the interplay of spin and valley relaxations in Si/SiGe quantum dots.
Our findings suggest that the electron spin dephases due to valley relaxation, and are in agreement with recent experimental studies [\href{http://www.nature.com/nnano/journal/v9/n9/abs/nnano.2014.153.html}{Nature Nanotechnology {\bf 9}, 666--670 (2014)}].
\end{abstract}

\maketitle

\section{INTRODUCTION}
Finding efficient ways to use electron spins in quantum dots (QDs) as quantum bits (qubits) has been an active field of research in condensed matter physics for many years [\onlinecite{Loss1}, \onlinecite{Petta1}, \onlinecite{Elzerman1}, \onlinecite{Veldhorst1}, \onlinecite{Yang1}, 
\onlinecite{Scarlino1}]. A necessary prerequisite for building qubits are long coherence times, long enough to 
allow for a large number of gate operations before the quantum-mechanical nature of the qubit is irreversibly lost [\onlinecite{DiVincenzo1}].
An electron spin confined in a semiconductor quantum dot loses its quantum phase coherence due to interactions with its noisy, solid state environment. Unavoidable interactions of the electron spin with surrounding charges and nuclear spins are common mechanisms 
that limit the coherence time of the electron spin $T_2^*$ to as little as nanoseconds in some structures [\onlinecite{Petta1}, \onlinecite{Hanson1}, \onlinecite{Coish1}, \onlinecite{Coish2}].

In natural silicon only $\approx$ 4.7 $\%$ of the atomic nuclei have a non-zero spin. Therefore, Si represents a logical candidate for
the implementation of spin qubits [\onlinecite{Veldhorst1}, \onlinecite{Morello1}, \onlinecite{Zajac1}, \onlinecite{Zwanenburg1}]. There are two implementation strategies for spin qubits in Si, using the nuclear spin of a phosphorus donor in Si [\onlinecite{Kane1}] and using spin states of an electron confined inside a Si quantum dot 
[\onlinecite{Vrijen1}, \onlinecite{Tahan2}, \onlinecite{Koiller1}, \onlinecite{Culcer1}].
Bulk silicon has six minima of the conduction band, known as valleys.
In a Si/SiGe quantum well, four out of six valley states are higher in energy due to strain at the Si/SiGe interface [\onlinecite{Schaeffler}]. The degeneracy of the remaining two valley states can be lifted by the confining potential in the $z$-direction 
[\onlinecite{Fusayoshi}, \onlinecite{Boykin}].

\begin{figure}[t!]
\centering
\includegraphics[height=6.0cm]{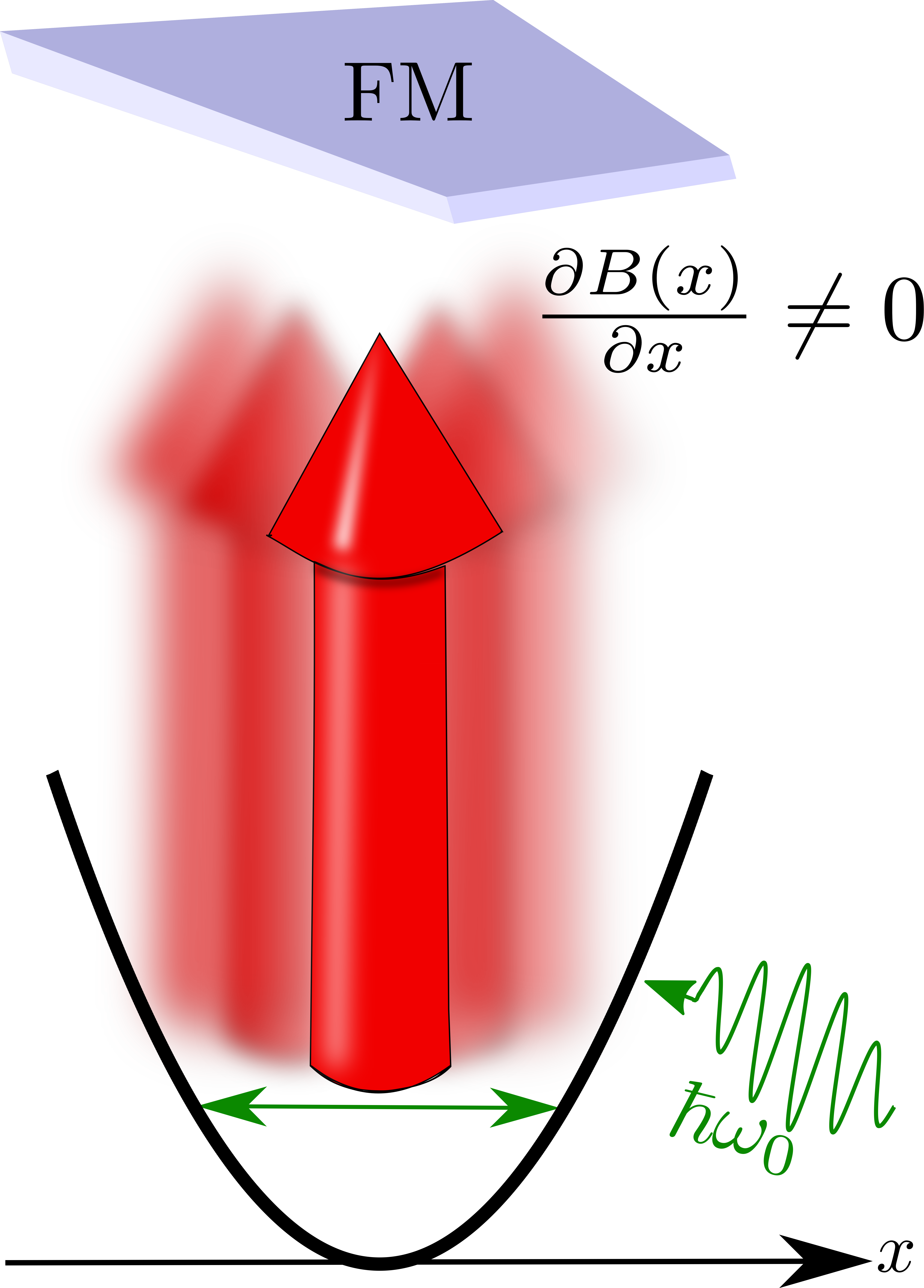}
\caption{(Color online) The control of the electron spin. A ferromagnet (FM) induces a magnetic field gradient in the $x$-direction. When microwave bursts are applied the electron experiences an effectively time dependent magnetic field in the direction of oscillation.
$\omega_0$ is the Larmor frequency of the microwaves.}
\label{Vis2}
\end{figure}

In this paper we study a situation in which a ferromagnet is embedded on top of the quantum dot, as shown in Fig. \ref{Vis2}. The in-plane component of the ferromagnet stray magnetic field, leads to the existence of a 
valley dependent $g$-factor, as predicted in the following theoretical study [\onlinecite{Tokura1}].
The goal of our theoretical study is to establish a quantitative relationship between valley dependent $g$-factors and the tilt of the Si/SiGe interface.
Consequently, the ferromagnet embedded on top of the quantum dot also leads to a valley dependent Rabi frequency [\onlinecite{Pioro1}]. The valley dependent $g$-factor causes the resonance condition to be different for the two valleys, and alongside with valley
dependent Rabi frequencies, leads to errors in controlling the electron spin state in one of the valleys.
Furthermore, when valley relaxation is present, a novel decoherence mechanism exists which cannot be reversed by a spin echo [\onlinecite{Kawakami1}].
If the electron is driven on resonance in one of the valleys, valley relaxation abruptly changes the resonance condition causing the electron spin to decohere. Another goal of this manuscript is to describe the reduction of electron spin coherence due to valley 
relaxation, by solving a Lindblad master equation. The presence of spin relaxation, alongside with valley relaxation, leads to a rich interplay of spin and valley relaxation, which is also described by solving a Lindblad master equation. 

This paper is organized as follows. In Section \ref{ValD} we quantitatively describe the valley dependent $g$-factor induced by an in-plane stray magnetic field.
We continue by discussing the existence of a valley dependent Rabi frequency in Section \ref{SecResNuc}.
Subsequently, in Section \ref{mod} we present the Hamiltonian and the Lindblad equation for the open-system dynamics of the electron spin and qualitatively and quantitatively describe the drop of the electron spin coherence caused by valley relaxation. 
In Section \ref{Res2} we discuss the interplay of valley and spin relaxations, before concluding in Section \ref{Con}.

\section{VALLEY DEPENDENT $\mathbf{g}$-FACTOR in $\text{{\bf Si/SiGe}}$ quantum dots}\label{ValD} %Intervalley scattering in Si Ferry Phys. %Rev. B 14 4 (1976) and New J. Phys. 15 125010

Bulk silicon has six effective minima of the conduction band named valleys. In a Si/SiGe quantum dot four of the valleys are lifted higher in energy by the presence of strain at the Si/SiGe interface and the two low energy valleys remain degenerate.
The degeneracy of the remaining two valleys is typically lifted by the confining potential in the $z$-direction [\onlinecite{Fusayoshi}, \onlinecite{Boykin}].

The Hamiltonian of a single electron confined in a Si/SiGe quantum dot in a magnetic field in the $z$-direction, and a magnetic field gradient in the $x$-direction is given by
\begin{equation}\label{eq:FullHam}
H=H_0+H_{\rm z}+H_{\rm FM}.
\end{equation}
Here, $H_0$ is the Hamiltonian of the single electron confined in a Si/SiGe quantum dot,
\begin{equation}\label{eq:ConfHam}
 H_0=\frac{p_z^2}{2m^*_z}+\frac{p_x^2+p_y^2}{2m^*_t}+V(x)V(y)V(z),
\end{equation}
where $p_i$ denotes the $i$-th component of the momentum operator, and $m^*_z$ the longitudinal electron mass (in the direction perpendicular to the Si/SiGe quantum well). Furthermore, $m^*_t$ is the transverse electron mass (in the plane of Si/SiGe 
quantum well) and $V(x)V(y)V(z)$ are confining potentials in the $x,y,z$ directions, respectively. The confining potentials in the $x$-direction and $y$-direction originate 
from the electrostatic confinement and are modeled with a harmonic oscillator potential $V(x)=m^*_t\omega_0^x x^2/2$, $V(y)=m^*_t\omega_0^y y^2/2$. 
The potential in the $z$-direction comes from the Si/SiGe quantum well and is modeled as a finite square well potential.
$H_{\rm z}$ is the Zeeman Hamiltonian 
\begin{equation}\label{eq:ZemHam}
 H_{\rm z}=g\mu_{\rm B} B_0S_z, 
\end{equation}
where $g$ is the electron $g$-factor, $\mu_{\rm B}$ is the Bohr magneton, $B_0$ is the total magnetic field (in the $z$-direction) and $S_z$ is the $z$ component of the electron spin operator.
Furthermore, $H_{\rm FM}$ is the Hamiltonian describing the stray field in the $x$-direction coming from the ferromagnet
\begin{equation}\label{eq:StrayHam}
 H_{\rm FM}=g\mu_{\rm B} B(x)S_x,
\end{equation}
where $S_x$ is the $x$ component of the electron spin operator and $B(x)$ is the $x$ component of the magnetic field coming from the ferromagnet $B(x)=B_x^0 x/a_{\rm B}$. Here $B_x^0$ is the strength of the slanting field, $x$ is the position operator 
and $a_{\rm B}=\sqrt{\hbar/m^*_t \omega_0^x}$ is the effective Bohr radius in the $x$-direction of the electron spin confined in a quantum dot, where $m^*_t$ is the transverse effective electron mass and $\omega_0^x$ is the confining potential in the $x$-direction.
An in-plane magnetic field gradient $B(x)$ modifies the Zeeman energy [\onlinecite{Tokura1}]. In our case the in-plane magnetic field gradient is caused by the ferromagnet embedded on top of the quantum dot (Fig. \ref{Vis2}). %cite PRL 96 047202 (2006) 
Neglecting the gradient in the $z$-direction is a good approximation when the total magnetic field (directed along $z$) is much larger than the $z$ component of the stray field. 

Proceeding similar to [\onlinecite{Tokura1}], energy levels of $H_0+H_{\rm z}$ are obtained as %cite PRL 96 047202 (2006)
\begin{equation}\label{eq:ener}
 E=E_n\pm E_{\rm z}/2.
\end{equation}
Here, $E_n$ is the confinement energy and $E_{\rm z}=g\mu_{\rm B} B_0$ is the electron Zeeman energy. A plus sign in Eq. (\ref{eq:ener}) stands for a spin-up state $|\!\uparrow\rangle$ and a minus sign for a spin-down state $|\!\downarrow \rangle$.

The first order energy correction coming from $H_{\rm FM}$ is zero because of the even parity of the ground state wavefunction of the linear harmonic oscillator (LHO) and odd parity of $H_{\rm FM}$. 
The second order energy correction coming from the magnetic field gradient term $H_{\rm FM}=g\mu_{\rm B}B(x)S_x$ yields

\begin{equation}\label{eq:pert}
E^{(2)}_{m_s}=-\frac{1}{4}\sum\limits_{n=1}^{\infty} \frac{M_n^2}{\Delta_n-2 m_s E_{\rm z}},
\end{equation}
where $m_s=\pm 1/2$ is the spin projection quantum number. The symbol $\Delta_n$ stands for the energy difference between the orbital ground state and the $n$-th state. 
Furthermore, $M_n$ is the matrix element between the ground state and the $n$-th orbital state of the LHO
\begin{equation}
 M_n =\langle\Psi_0\!\uparrow\!|H_{\rm FM}|\!\downarrow\!\Psi_n\rangle=\frac{g \mu_{\rm B} B_x^0}{2 a_{\rm B}} \langle \Psi_0|x|\Psi_n \rangle,
\end{equation}
where $\Psi_0$ is the ground state LHO wavefunction and $\Psi_n$ is the LHO wavefunction of the $n$-th excited state. 
Because
\begin{equation}\label{eq:cas}
\langle  \Psi_0 |x|\Psi_n\rangle=\frac{1}{\sqrt{2}}a_{\rm B} \delta_{n,1}
\end{equation}
and because $S_x$ couples only states with different spin projections $m_s$, for an electron in the ground orbital state the sum in Eq. (\ref{eq:pert}) 
is substituted by a single term with a matrix element
\begin{equation}\label{eq:Mel}
M_1 = \frac{g\mu_{\rm B} B^0_x}{2\sqrt{2}}.
\end{equation}
Therefore, the slanting magnetic field in the $x$-direction corrects the ground state energy of the electron

\begin{equation}\label{eq:pert2}
 E^{(2)}_{m_{s}}=-\frac{1}{4}\frac{M^2_1}{\Delta-2 m_s E_{\rm z}},
\end{equation}
where $\Delta=\hbar \omega_0^x$ is the orbital splitting.

In the presence of valley-orbit mixing the orbital splitting $\Delta_{v,\bar{v}}$ is valley dependent [\onlinecite{Friesen1}, \onlinecite{Gamble1}]. 
This yields a valley dependent energy correction due to the slanting magnetic field Eq. (\ref{eq:pert2}), and therefore a correction to the electron $g$-factor which depends on the valley state,  
\begin{equation}\label{eq:gValley}
 g_j=\frac{g}{E_{\rm z}}\bigg(E_{\rm z}+E_{\uparrow,j}^{(2)}-E_{\downarrow,j}^{(2)}\bigg)=g\Bigg[1-\frac{1}{2}\frac{M^2_1}{\Delta_{j}^2-E_{\rm z}^2}\Bigg].
\end{equation}
\begin{figure}[t!]
\centering
\includegraphics[width=0.45\textwidth]{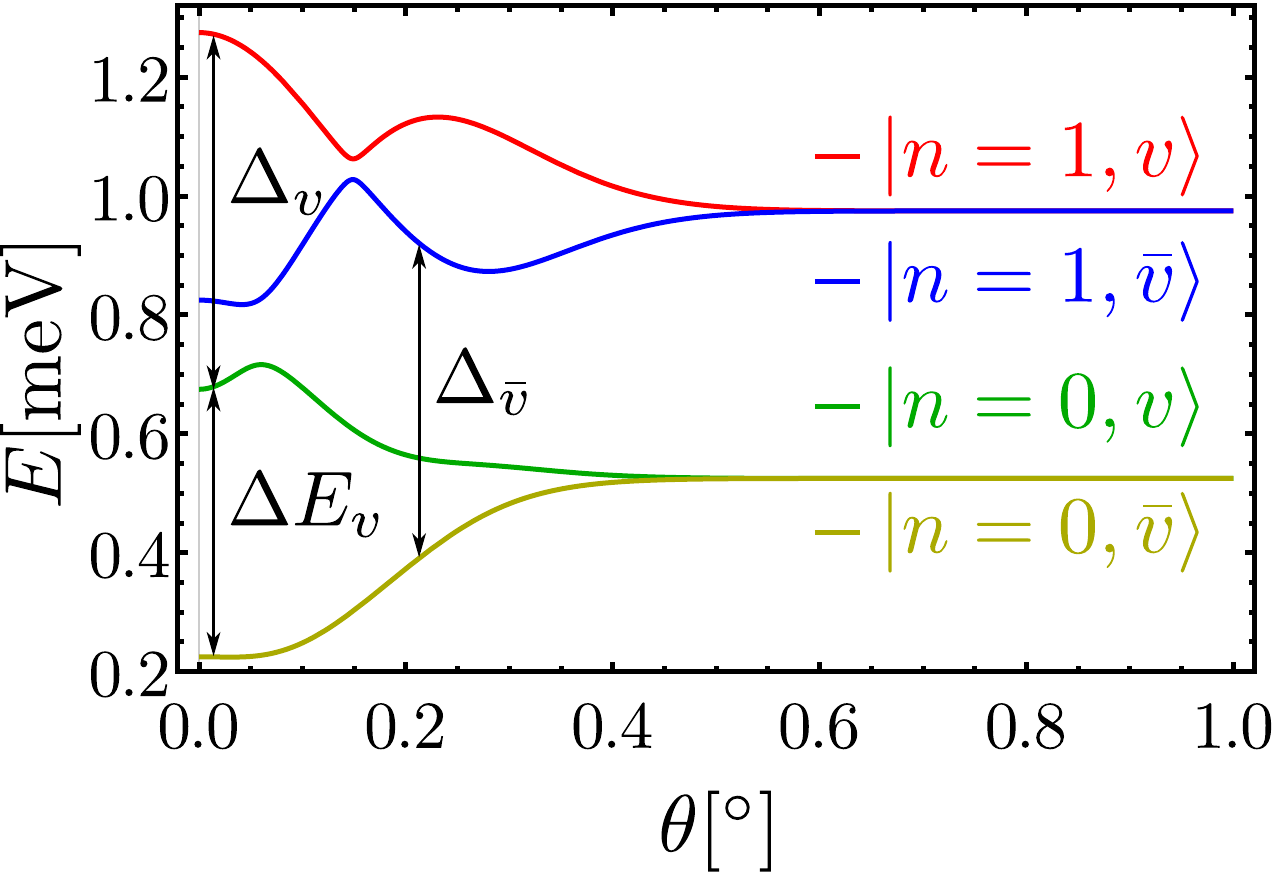}
\caption{(Color online) The lowest four energy states as a function of the effective miscut angle $\theta$. $\Delta E_v$ is the ground state valley splitting, and $\Delta_v$ and $\Delta_{\bar{v}}$ are orbital splittings in the $v$ and $\bar{v}$ valleys.
The parameters of the plot are $\hbar \omega_0^x=450\text{ $\mu$eV}$, $v_v \xi^2(z_0)=300\text{ $\mu$eV}$, $k_0=2 \pi\cdot 0.82/a$, where $a=5.431\text{ \AA}$ is the lattice constant of Si, and $m_t^*=0.19 m_e$.
It should be noted that due to valley-orbit mixing the orbital quantum numbers $n=0,1$ and valley quantum numbers $v=\pm 1$ 
are not good quantum numbers anymore.}
\label{spectrum}
\end{figure}

\begin{figure*}[t!]
\includegraphics[width=\textwidth]{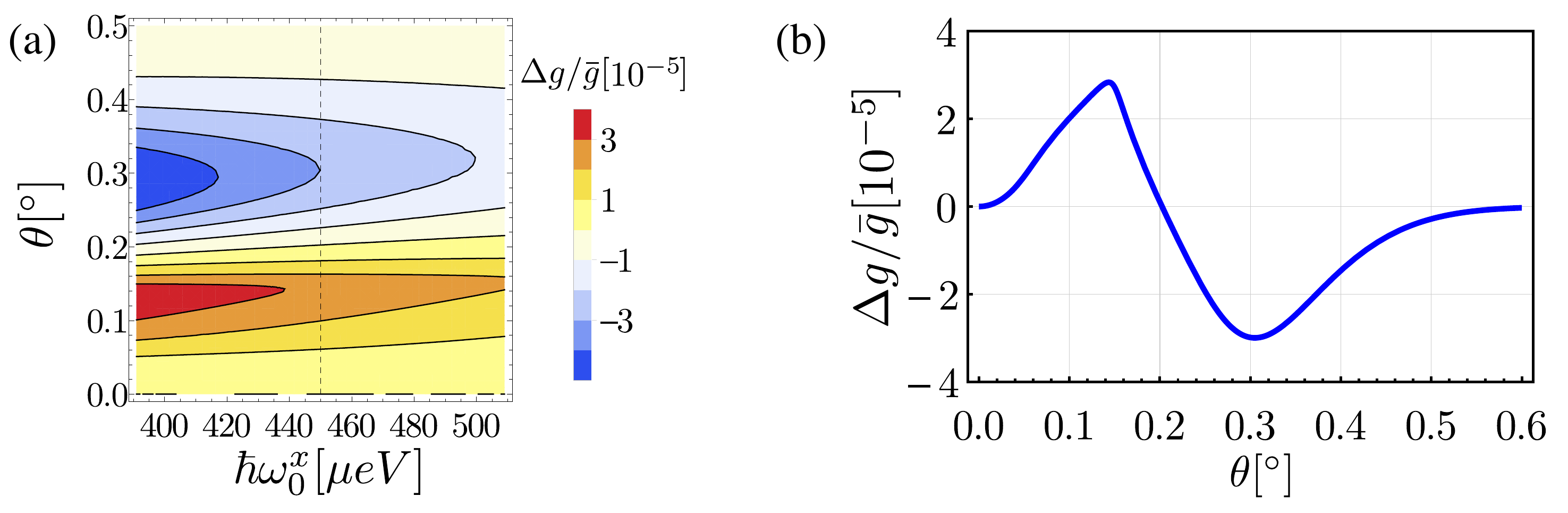}
\caption{(Color online) (a) The average difference of valley $g$-factors as a function of the effective tilt angle $\theta$ and the confinement energy $\hbar \omega_0^x$.
(b) The average difference of valley $g$-factors for the value of the single orbital spacing $ \hbar \omega_0^x=450\text{ $\mu$eV}$ (see dashed line in (a)) [\onlinecite{Kawakami1}].
The parameters of the plots are the following, 
$v_v \xi^2(z_0)=300\text{ $\mu$eV}$, $m_t^*=0.19 m_e$ and $k_0=2 \pi \cdot 0.82/a$, where $a=5.431\text{ \AA}$ is the lattice constant of Si, $B_0^x=3.4\text{ mT/nm}$, $B_z=0.75\text{ T}$, 
the $z$-component of the magnetic field of the ferromagnet $B_z^{\rm FM}=-0.12\text{ T},$ and the height of the Si quantum well is $z_0=12\text{ nm}$.}
\label{gfact}
\end{figure*}
Here, $ g_{j}$ is the $g$-factor corresponding to two valley states $j=\{v,\;\bar{v}\}$ and $\Delta_{j}$ is the valley dependent orbital level spacing corresponding to $j$-th valley state. The average difference of valley $g$-factors $\Delta g/\bar{g}$
is defined as

\begin{equation}\label{eq:deltagB}
\frac{\Delta g}{\bar{g}}=2\frac{g_v-g_{\bar{v}}}{g_v+g_{\bar{v}}}.
\end{equation}
Inserting equation Eq. (\ref{eq:gValley}) into Eq. (\ref{eq:deltagB}) we obtain

\begin{equation}\label{eq:deltag}
\frac{\Delta g}{\bar{g}}= \frac{2M_1^2(\Delta^2_{\bar{v}}-\Delta^2_{v})}{(\Delta^2_{v}-E_{\rm z}^2)(\Delta^2_{\bar{v}}-E_{\rm z}^2)-M^2_1(\Delta^2_{v}+\Delta^2_{\bar{v}}-2E_{\rm z}^2)/4}.
\end{equation}
Here, $\Delta_{j}$ is the energy difference between the orbital ground state and the first excited orbital state in the $j$-th valley. Furthermore, $E_{\rm z}$ is the Zeeman energy, and $M_1$ is the matrix element between the orbital ground
state and the first excited orbital state coming from the slanting field Eq. (\ref{eq:Mel}).

Valley-orbit mixing $\Delta_{v}-\Delta_{\bar{v}}\neq 0$ occurs due to miscuts of the Si/SiGe quantum well [\onlinecite{Zwanenburg1}, \onlinecite{Ando1}]. 
The valley coupling can be described by a $\delta$ function [\onlinecite{Goswami1}, \onlinecite{Friesen2}]
\begin{equation}\label{eq:vv}
 V_v({\bf r})=v_v\delta(z-z_0+\theta x).
\end{equation}
Here, $z_0$ is the position of the SiGe interface, the miscut is usually between $0^{\circ}\le\theta\le 2^{\circ}$, so it is safe to approximate $\tan(\theta) \approx \sin(\theta)\approx \theta $. Furthermore, $v_v$ is the valley coupling strength.
We have further assumed for simplicity that the miscut occurs in the $x$-direction, and therefore the valley coupling operator Eq. (\ref{eq:vv}) does not depend on the $y$ component.

As the wavefunction is closest to the top interface only one delta function potential is present in the theory. It should be noted that the atomistic details of the valley splitting are included via the valley coupling potential Eq. (\ref{eq:vv}) and therefore
the periodic parts of the Bloch wavefunctions do not play a role in our calculations and can be omitted. 
Treating valley coupling as a perturbation the general formula for matrix elements of the valley coupling operator Eq. (\ref{eq:vv})

\begin{multline}\label{eq:Mat2}
  \langle n',\bar{v}| V_v({\bf r})| n,v \rangle=v_v \xi^2(z_0)e^{2ivk_0z_0}\times \\
\times \int_{-\infty}^{\infty} e^{-2 i v k_0 x\theta}\Psi_{n'}(x)\Psi_{n}^*(x) dx.
\end{multline}
Assuming that the wavefunctions $\Psi_n$ are those of the LHO the diagonal elements of the valley coupling operator Eq. (\ref{eq:vv}) have the following form

\begin{equation}\label{eq:Mat1}
 \langle n,v|V_v({\bf r})| n,v \rangle = v_v\xi^2(z_0),
\end{equation}
where $n$ is the orbital quantum number corresponding to the wavefunction in the $x$-direction, $v$ is the valley quantum number, $\xi(z_0)$ is the ground state electron wavefunction in the $z$-direction and $z_0$ is the position of the Si/SiGe interface. 
Due to the fact that the confinement in the $z$-directions comes from a sharp Si/SiGe interface, the orbital level spacing in the $z$-direction is large, so we assume that the system is always in the ground state in the $z$-direction.
The off-diagonal matrix elements of the lowest two orbital states of the valley coupling operator Eq. (\ref{eq:vv}) have the following form

\begin{gather}
\langle 0,\bar{v}|V_v({\bf r})|1,v\rangle=-i\sqrt{2}v_v\xi^2(z_0) k_0\theta a_{\rm B} e^{2ik_0z_0} e^{-k_0^2\theta^2 a_{\rm B}^2},\nonumber \\
\langle 0,\bar{v}|V_v({\bf r})|0,v\rangle=v_v\xi^2(z_0) e^{2ik_0z_0} e^{-k_0^2\theta^2 a_{\rm B}^2},\nonumber \\
\langle 1,\bar{v}|V_v({\bf r})|1,v\rangle=v_v\xi^2(z_0)(1-2k_0^2\theta^2 a_{\rm B}^2)e^{2ik_0z_0} e^{-k_0^2\theta^2 a_{\rm B}^2}.\label{eq:OffDa1}
\end{gather}
Here $k_0$ is the reciprocal lattice constant of Si, $z_0$ is the position of the Si/SiGe interface, $\theta$ is the effective tilt angle, and $a_{\rm B}$ is the effective Bohr radius in the $x$-direction.
Constraining the discussion on the lowest two orbital states, and diagonalizing the matrix constituted of elements from Eq. (\ref{eq:Mat1}) and Eq. (\ref{eq:OffDa1}) we obtain the mixed valley-orbit eigenspectrum Fig. \ref{spectrum}
 (and therefore $\Delta_v$ and $\Delta_{\bar{v}}$).
 
\begin{figure}[t!]
\centering
\includegraphics[width=0.45\textwidth]{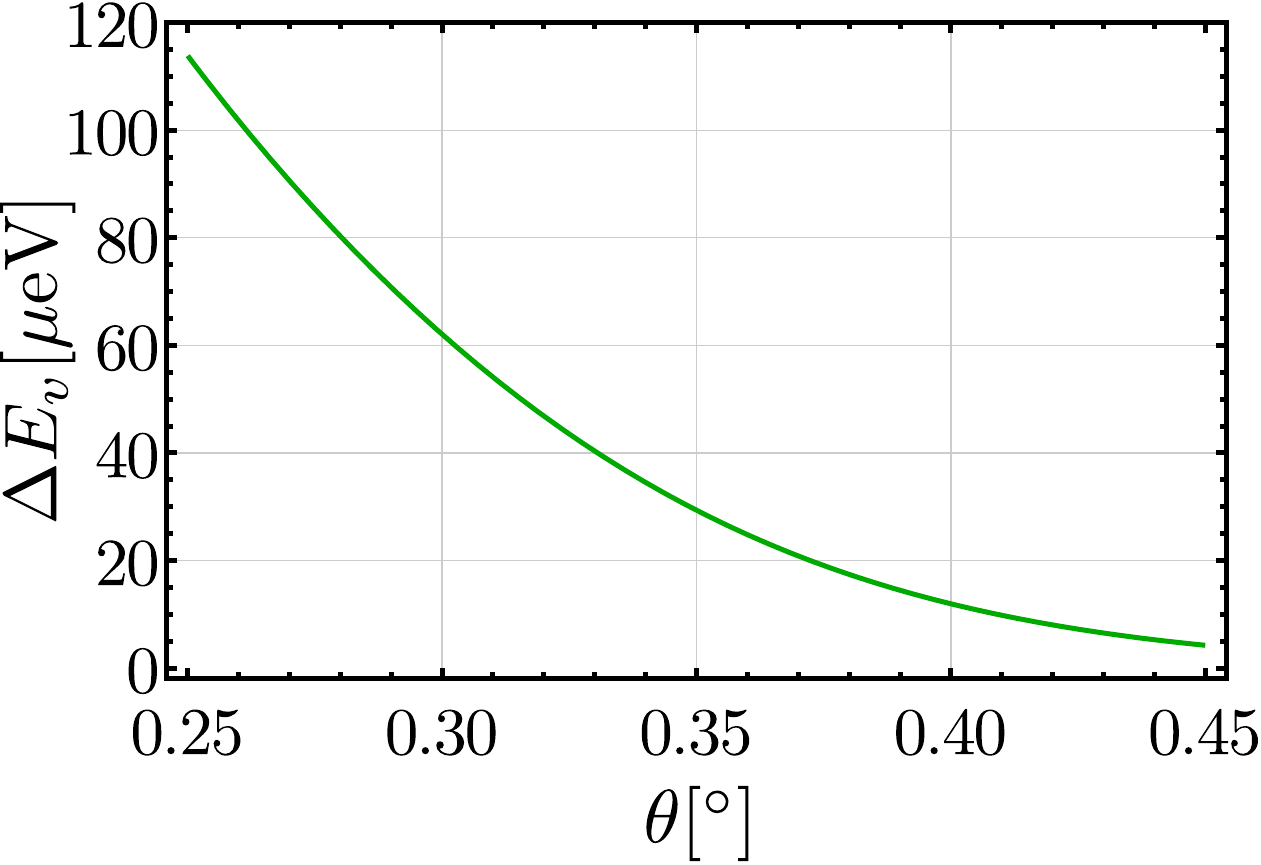}
\caption{(Color online) Ground state valley splitting $\Delta E_v$ as a function of the effective tilt angle $\theta$. The parameters of the plot are, $\hbar \omega_0^x=450\text{ $\mu$eV}$, 
$v_v \xi^2(z_0)=300\text{ $\mu$eV}$, $m_t^*=0.19 m_e$ and $k_0=2 \pi 0.82/a$, where $a=5.431\text{ \AA}$ is the lattice constant of Si, $B_0^x=3.5\text{ mT/nm}$, and the
size of the Si quantum well is $z_0=12\text{ nm}$.}
\label{valleysplitting}
\end{figure}

Constraining the discussion again on the lowest two orbital states, diagonalizing the matrix constituted of elements from Eq. (\ref{eq:Mat1}) and Eq. (\ref{eq:OffDa1}), and inserting the result of the diagonalization into Eq. (\ref{eq:deltag}) we obtain
the average difference of electron $g$-factors as a function of the confining energy $\hbar\omega_0^x$ and the effective tilt angle $\theta$ (Fig. \ref{gfact}).
In Fig. \ref{gfact} (a) we see that for $\theta \approx 0.2^\circ$ the average difference of valley dependent $g$-factors goes to zero due to the fact that for this particular value of the effective tilt angle $\Delta_{v}\approx\Delta_{\bar{v}}$. 
Recent experimental studies [\onlinecite{Kawakami1}] yield an absolute average difference of valley $g$-factors of $|\Delta g/\bar{g}|=1.5\cdot10^{-4}$ and predict an absolute average difference of $g$-factors of
$|\Delta g/\bar{g}|=3\cdot10^{-5}$, given the single orbital spacing $\hbar \omega_0^x=450\text{ $\mu$eV}$. In our calculations $|\Delta g/\bar{g}|=3\cdot10^{-5}$ corresponds to the values $\theta\approx 0.15^\circ$ or 
$\theta\approx 0.3^\circ$ for $\hbar \omega_0^x=450\text{$\mu$eV}$. When we plot the difference of the lowest two eigenvalues Fig. \ref{valleysplitting},
we see that the valley splitting corresponding to $\theta \approx 0.3 ^{\circ}$ is $E_v\approx 60\text{ $\mu$eV}$, in agreement with the typical value for quantum dots $\Delta E_v\sim0.1\text{ meV}$ [\onlinecite{Zwanenburg1}].

\section{VALLEY DEPENDENT RABI FREQUENCY}\label{SecResNuc}

When controlling the electron spin by oscillating it inside an in-plane magnetic gradient the Rabi frequency is calculated with the following formula [\onlinecite{Pioro1}]
\begin{equation}
\Omega=\frac{g\mu_{\rm B}}{2\hbar}eE_{\rm gate}\left|\frac{\partial B(x)}{\partial x}\right|\frac{a_{\rm B}^2}{\Delta}.
\end{equation}
Here, $E_{\rm gate}$ is the electric field of the gate, $B(x)$ is the in-plane magnetic field, $a_{\rm B}=\sqrt{\hbar/\omega_0 m^*_t}$ is the effective Bohr radius of the electron and $\Delta$ is orbital level spacing. As seen in Section \ref{ValD}, the $g$-factors
corresponding to different valleys only differ by $\approx 10^{-3}$ relative to their values, so throughout this Section it is safe to assume that $g_v=g_{\bar{v}}=g=2$.
If the valley and orbit degree of freedom mix
(due to, e.g., Si/SiGe interface miscut) the orbital level spacing $\Delta_{v,\bar{v}}$, and therefore the Rabi frequency $\Omega_{v,\bar{v}}$ become valley dependent, with a relative difference of Rabi frequencies 

\begin{equation}\label{eq:DeltaOmega}
\frac{\Delta \Omega}{\bar{\Omega}}=2\frac{\Delta_{\bar{v}}-\Delta_v}{\Delta_{\bar{v}}+\Delta_v}.
\end{equation}

\begin{figure}[t!]
\centering
\includegraphics[width=0.43\textwidth]{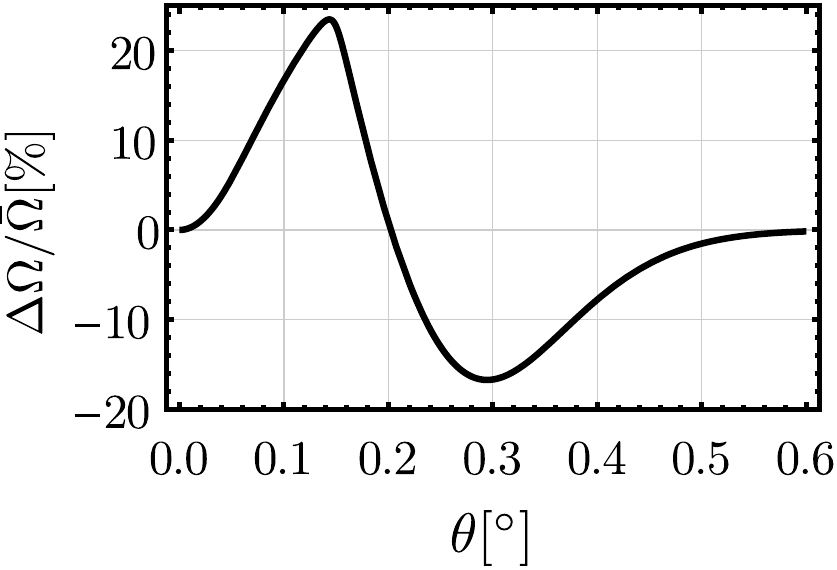}
\caption{(Color online) Average difference of Rabi frequencies $\Delta\Omega/\bar{\Omega}$ as a function of the effective tilt angle $\theta$. The parameters of the plot are, $\hbar \omega_0^x=450\text{ $\mu$eV}$, 
$v_v \xi^2(z_0)=300\text{ $\mu$eV}$, $m_t^*=0.19 m_e$ and $k_0=2\cdot\pi 0.82/a$, where $a=5.431\text{ \AA}$ is the lattice constant of Si, $B_0^x=3.5\text{ mT/nm}$, $B_z=0.75\text{ T}$, and the
height of the Si quantum well is $z_0=12\text{ nm}$.}
\label{DelOm}
\end{figure}

By diagonalizing a matrix whose terms are constituted from Eq. (\ref{eq:Mat1}) and Eq. (\ref{eq:Mat2}) and then inserting the result into Eq. (\ref{eq:DeltaOmega}) we obtain the average difference of Rabi frequencies as a function of the effective tilt angle $\theta$ (Fig. \ref{DelOm}).
A $50\%$ absolute average difference of valley Rabi frequencies has been measured in a recent experimental study [\onlinecite{Kawakami1}]. In our case the maximum $\Delta \Omega/\bar{\Omega}=25\%$, which corresponds to a value of the effective tilt angle $\theta \approx 0.15^{\circ}$ (see Fig. \ref{DelOm}). 
The discrepancy between our theory and the experiment may be due the fact that the product of valley coupling strength and the square of the wavefunction at the position of the Si/SiGe interface $v_v\xi^2(z_0)$, is a free parameter.
$v_v$ depends on the abundance of Ge $x$ in the Si/Si$_x$Ge$_{1-x}$ quantum well and can be estimated from tight binding theories [\onlinecite{Friesen2}]. On the other hand, 
$\xi^2(z_0)$ depends on the thickness of the Si layer and the exact type of the confinement in the Si/SiGe quantum well.

\begin{figure}[t!]
\centering
\includegraphics[height=5.3cm]{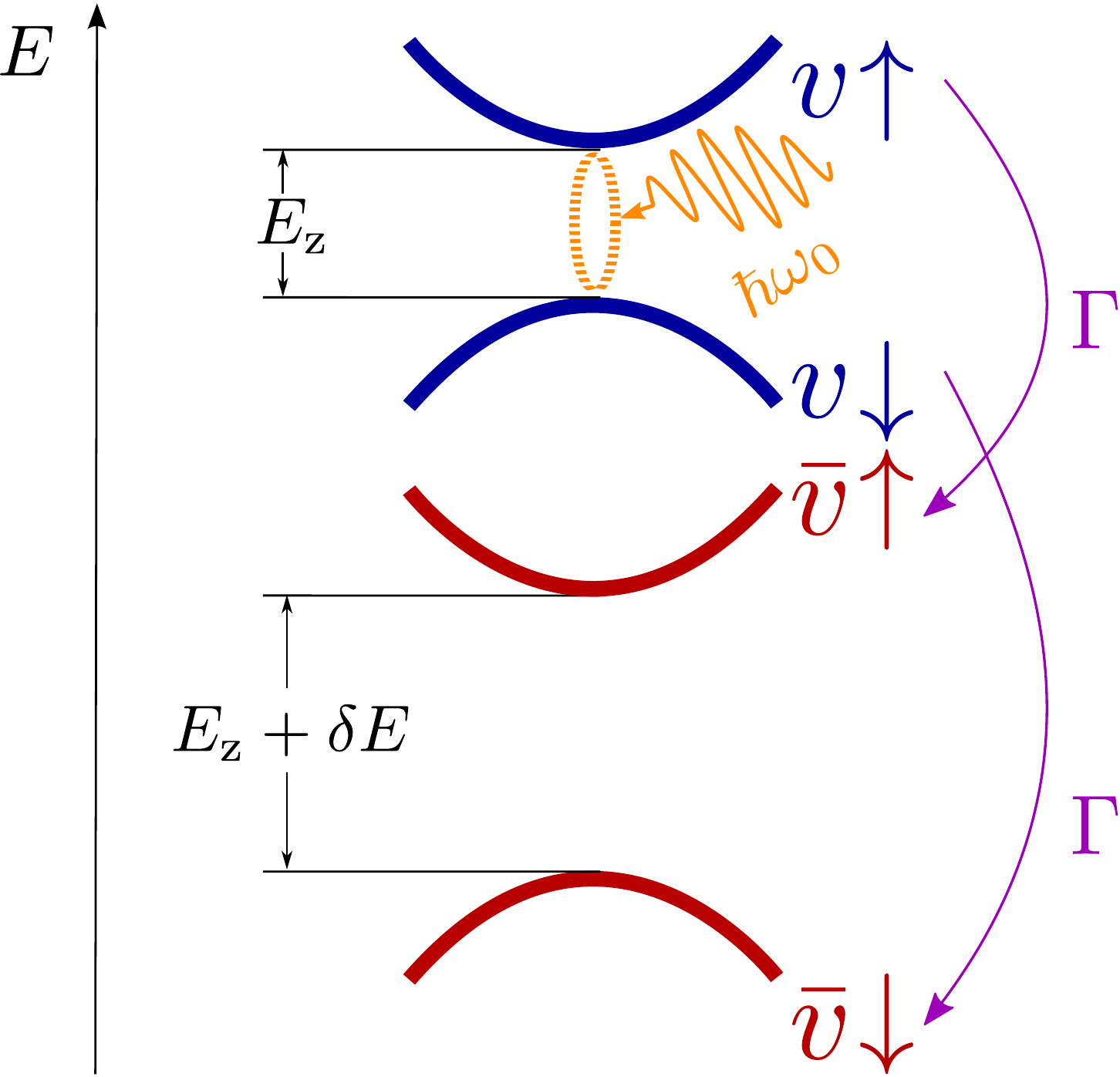}
\caption{(Color online) Visualizing a valley dependent $g$-factor. $\Gamma$ is the valley relaxation rate, $\delta E=(g_v-g_{\bar{v}})\mu_{\rm B} B_z$ is the difference of valley Zeeman energies, $E_z$ is the Zeeman energy of the confined electron and $\omega_0$ is the Larmor frequency.}
\label{Vis3}
\end{figure}

\section{MODELING THE DECOHERENCE}\label{mod}

We model a situation in which an electron spin is confined in a Si/SiGe quantum dot with a ferromagnet embedded 
on top of the quantum dot [\onlinecite{Tokura1}], inducing a stray magnetic field as shown in Fig. \ref{Vis2}.
All-electrical two-axis control of single electron spin states is achieved by oscillating the electron in real space with microwave bursts [\onlinecite{Tokura1}, \onlinecite{Wu1}] (Fig. \ref{Vis2}).
As the electron oscillates in real space it experiences a periodic, time-dependent, magnetic field.

The free evolution of the electron spin is described by the following Hamiltonian
\begin{equation}\label{eq:Ham0}
 H_0=\sum\limits_{\sigma=\downarrow,\uparrow}\sum\limits_{j=v,\bar{v}}E_{\sigma j}c^{\dagger}_{\sigma j}c_{\sigma j}.
\end{equation}
Microwave induced oscillations of the electron in real space, combined with the stray field of the ferromagnet, alter the state of the electron spin, while leaving the valley degree of freedom unchanged. Coupling to the microwave field is described by
\begin{equation}
 H'(t)=\sum\limits_{j=v,\bar{v}}\hbar\Omega_j\cos{(\omega t)}(c^{\dagger}_{\downarrow j}c_{\uparrow j}+H.c.).
\end{equation}
Applying the rotating wave approximation to the Hamiltonian $H_0+H'(t)$, we obtain the time-independent Hamiltonian in the rotating frame,

\begin{widetext}
\begin{equation}\label{eq:Ham}
H=\frac{1}{2}
\begin{pmatrix}
E_{\rm z}-\hbar \omega_0&& \hbar \Omega_v&&0&&0\\
\hbar \Omega_v&&-E_{\rm z}+\hbar \omega_0&&0&&0\\
0&&0&&E_{\rm z}+\delta E-\hbar \omega_0&&\hbar \Omega_{\bar{v}}\\
0&&0&&\hbar \Omega_{\bar{v}}&&-E_{\rm z}-\delta E+\hbar \omega_0
\end{pmatrix},
\end{equation}
\end{widetext}
in the $\{v \uparrow, v \downarrow, \bar{v} \uparrow, \bar{v} \downarrow\}$ basis, where the $\{v$, $\bar{v}\}$ represent valley states, and $\{\uparrow,$ $\downarrow\}$ stand for spin states. 
$E_{\sigma j}$ is the energy of the $j$-th valley state with spin $\sigma$,
$c_{\sigma j}$ and $c_{\sigma j}^{\dagger}$ are electron creation and annihilation operators. Furthermore, $E_{\rm z}$ is the Zeeman energy
of the electron, $\omega_0$ is the Larmor frequency, $\Omega_{v,\bar{v}}$ is the valley dependent Rabi frequency and $\delta E=(g_v-g_{\bar{v}})\mu_{\rm B} B_z$ is the difference of valley Zeeman energies Fig. \ref{Vis3}.

The goal of our study is to model the influence of valley relaxation on electron spin coherence. An electron is initialized in the $|\!\downarrow\rangle$ state with valley injection probabilities $P^0_v=0.7$, $P^0_{\bar{v}}=0.3$.
We model a spin echo experiment, where first a $\pi/2$ pulse is applied, followed by a free (undriven) evolution of a duration $t/2$. Afterwards, a $\pi$ pulse is applied followed by another free evolution of a duration $t/2$ and another $\pi/2$ pulse. 

The valley relaxation is assumed to occur only during the free evolution stage (as the duration of the free evolution stage $t$ is much larger than the duration of $\pi$ pulses), and is modeled with a Lindblad equation

\begin{multline}\label{eq:Lind}
\hspace{-3mm}{\dot{\rho}}=-\frac{i}{\hbar}[H_0,\rho]+\frac{1}{2}\Gamma(2 L^{\dagger}\rho L-L L^{\dagger}\rho-\rho L L^{\dagger})=\mathcal{L}\rho.
\end{multline}
Here, $\mathcal{L}$ is the Lindblad superoperator. Furthermore, $\Gamma$ is the valley relaxation rate, and $ L^{\dagger}=|v\rangle\langle\bar{v}|$ and $L=|\bar{v}\rangle\langle v|$
are Lindblad inter-valley dissipation operators.

We use the echo envelope function as a measure of the electron spin coherence. 
In order to be able to derive the echo envelope function, instead of using the Lindblad equation in the mentioned form Eq. (\ref{eq:Lind}) we use the Lindblad equation in superoperator form
\begin{equation}\label{eq:Sup}
\rho(t)=e^{\mathcal{L}t}\rho(0).
\end{equation}
Writing the Lindblad equation in the superoperator form allows us to include a sequence of $\pi /2-\pi-\pi/2$ pulses, around the $x$ axis, with inter-valley scattering occurring in the free evolution stage in the following way
\begin{equation}\label{eq:SupEch}
\rho(t)=R_x(\pi/2)e^{\mathcal{L}t/2}R_x(\pi)e^{\mathcal{L}t/2}R_x(\pi/2)\rho(0).
\end{equation}
Here, $R_x(\beta)$ rotates the spin $\rho(t)$ about an angle $\beta$ around the $x$ axis on the Bloch sphere. The $\pi$ and $\pi/2$ pulses are achieved by applying microwave pulses with a duration $\pi/\Omega_v$ and $\pi/2\Omega_v$, 
described by time evolution operators $R_x(\pi)=\exp{(-iH\pi/\hbar\Omega_v)}$ and $R_x(\pi/2)=\exp{(-iH\pi/2\hbar\Omega_v)}$, with $H$ being given by Eq. (\ref{eq:Ham}).

Finally, we obtain the echo envelope function, the probability that the electron changes spin to the $|\!\uparrow \rangle$ state after a total time of a free evolution $t$, when being subjected to a sequence of $\pi/2-\pi-\pi/2$ pulses
\begin{equation}\label{eq:Proj}
P_\uparrow=\sum\limits_{j=v,\bar{v}}{\rm Tr}(M_{\uparrow}^j \rho(t)).
\end{equation}
Here the $M_{\uparrow}^j$ are spin-up projection operators corresponding to $j$-th valley state.

\begin{figure}[t!]
\centering
\includegraphics[width=0.48\textwidth]{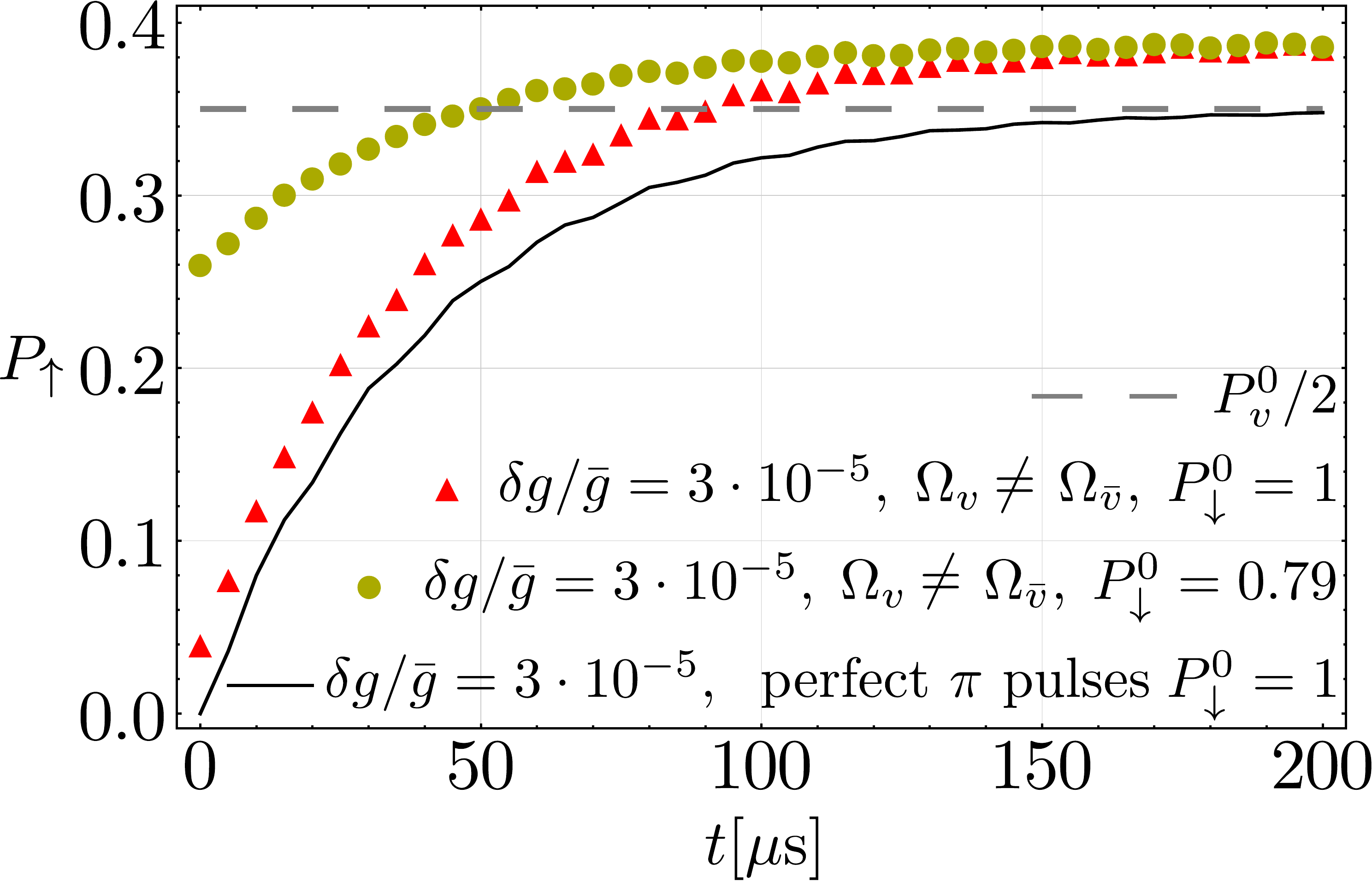}
\caption{(Color online) Probability that the echo sequence yields the electron $|\!\uparrow\rangle$ state. Red triangles and black line are a result of a simulation with injection probabilities $P_{v,\downarrow}^0=0.7$, $P_{\bar{v},\downarrow}^0=0.3$. Yellow disks are a result of a
simulation with injection probabilities $P_{v,\downarrow}^0=0.49$, $P_{v,\uparrow}^0=0.21$ and $P_{\bar{v},\downarrow}^0=0.3$.
The parameters of the plot are, the external magnetic field $B_z=0.75\text{ T}$,
 the $z$-component of the magnetic field of the ferromagnet $B_z^{\rm FM}=-0.12\text{ T}$, valley dependent Rabi frequencies corresponding to the miscut angle $\theta\approx 0.3^{\circ}$, $\Omega_v=2\pi\cdot3.1\text{ MHz}$, $\Omega_{\bar{v}}=2\pi\cdot3.7\text{ MHz}$,
 within the values suggested in a recent experimental study [\onlinecite{Kawakami1}].} 
\label{Res1}
\end{figure}

While the electron $g$-factor is valley dependent, the $\pi$ and $\pi/2$ pulses are still assumed perfect (see Fig. \ref{Res1}, black line), and
a valley relaxation event abruptly changes the resonance condition for $\delta E$ (see Eq. (\ref{eq:Ham})).
After the initial perfect $\pi/2$ pulse, in one half of the cases of inter-valley relaxation from $|v\rangle$ to $|\bar{v}\rangle$ the electron spin is in $|\!\uparrow\rangle$ state.
This is why the increase of the probability $P_\uparrow$, originating from valley relaxation, saturates at $P_v^0/2$ (see Fig. \ref{Res1}, gray dashed line).

The red triangles in Fig. \ref{Res1} represent the result of our simulation when the $g$-factors are valley dependent throughout the free evolution stage and the $\pi$ and $\pi/2$ pulses are imperfect in one of the valleys due to valley dependent $g$-factors and Rabi frequencies. 
After the imperfect initial $\pi/2$ pulse, the electron spin is not perpendicular to the magnetic field yielding rotations around the quantization axis
 with a frequency proportional to the Zeeman energy $g_{\bar{v}}\mu_{\rm B}(B_z+B_z^{\rm FM})/h$, where $B_z$ is the external magnetic field and $B_z^{\rm FM}$ is the $z$-component of the magnetic field of the ferromagnet. 
For $B_z=0.75\text{ T}$ and $B_z^{\rm FM}=-0.12\text{ T}$ this oscillations take place on a $\sim50\text{ ps}$ timescale, 
with the amplitude of the oscillations being given by the valley dependent Rabi frequencies $\Omega_v$ and $\Omega_{\bar{v}}$ and $g$-factors $g_v$ and $g_{\bar{v}}$. 
Therefore, the probability $P_\uparrow$ is very sensitive to the duration of the free evolution stage.
Due to the fact that the results of a recent experimental study [\onlinecite{Kawakami1}] represent an average over 150-1000 experimental outcomes, our results (red triangles and yellow discs, Fig. \ref{Res1}) represent an average over 1000 outcomes, 
randomly sampled from a $5\text{ ns}$ interval. When we compare the increase in probability due to valley relaxation (black line, Fig. \ref{Res1}) and the additional effect of imperfect $\pi$ and $\pi/2$ pulses
we see that imperfect rotations provide an additional mechanism that further increases $P_{\uparrow}$. 

A recent experimental study [\onlinecite{Kawakami1}] shows a fast initial increase by 0.25 of the probability $P_\uparrow$. Our model explains an initial increase of $P_\uparrow$ by a few $\%$ due to the averaging of the amplitude of 1000 randomly selected 
data points of the  $P_\uparrow$ oscillations close to $t=0$, occurring due to imperfect $\pi$ and $\pi/2$ pulses alongside with rotations around the $z$-axis in the free evolution stage. 
One possible explanation for the remaining discrepancy between the experimental findings 
and theory may be the initialization to the $|\!\downarrow\rangle$ state with a $\approx0.79$ fidelity (yellow disks, Fig. \ref{Res1}).
\section{INTERPLAY BETWEEN VALLEY AND SPIN RELAXATION}\label{Res2}

In Si quantum dots orbital relaxation happens on the $10^{-12}-10^{-7}\text{ s}$ time scale, spin relaxation on the $10^{-6}-1\text{ s}$ scale, and valley relaxation is somewhere between the two values [\onlinecite{Tahan1}].  %cite Charlie Tahan %GammaT1=10 kHz
In order to include spin relaxation processes we add an additional term to our Lindblad equation Eq. (\ref{eq:Lind}),
\begin{multline}\label{eq:SupEch2}
{\dot{\rho}}=-\frac{i}{\hbar}[H_0,\rho]+\frac{1}{2}\Gamma(2 L^{\dagger}\rho L-L L^{\dagger}\rho-\rho L L^{\dagger})\\
            +\frac{1}{2}\gamma(2 \sigma_{+}\rho \sigma_{-}-\sigma_{-}\sigma_{+}\rho-\rho \sigma_{-}\sigma_{+})=\mathcal{L'}\rho,
\end{multline}
where $\mathcal{L'}$ is the Lindblad superoperator. Other than the terms introduced in Eq. (\ref{eq:Lind}), the newly introduced terms are the spin relaxation rates $\gamma$ and two new Lindblad dissipation operators related to spin relaxation 
$\sigma_{+}={|\!\uparrow\rangle\langle \downarrow\!|}$ and $\sigma_{-}={|\!\downarrow\rangle\langle\uparrow\!|}.$
\begin{figure}[t!]
\centering
\includegraphics[width=0.48\textwidth]{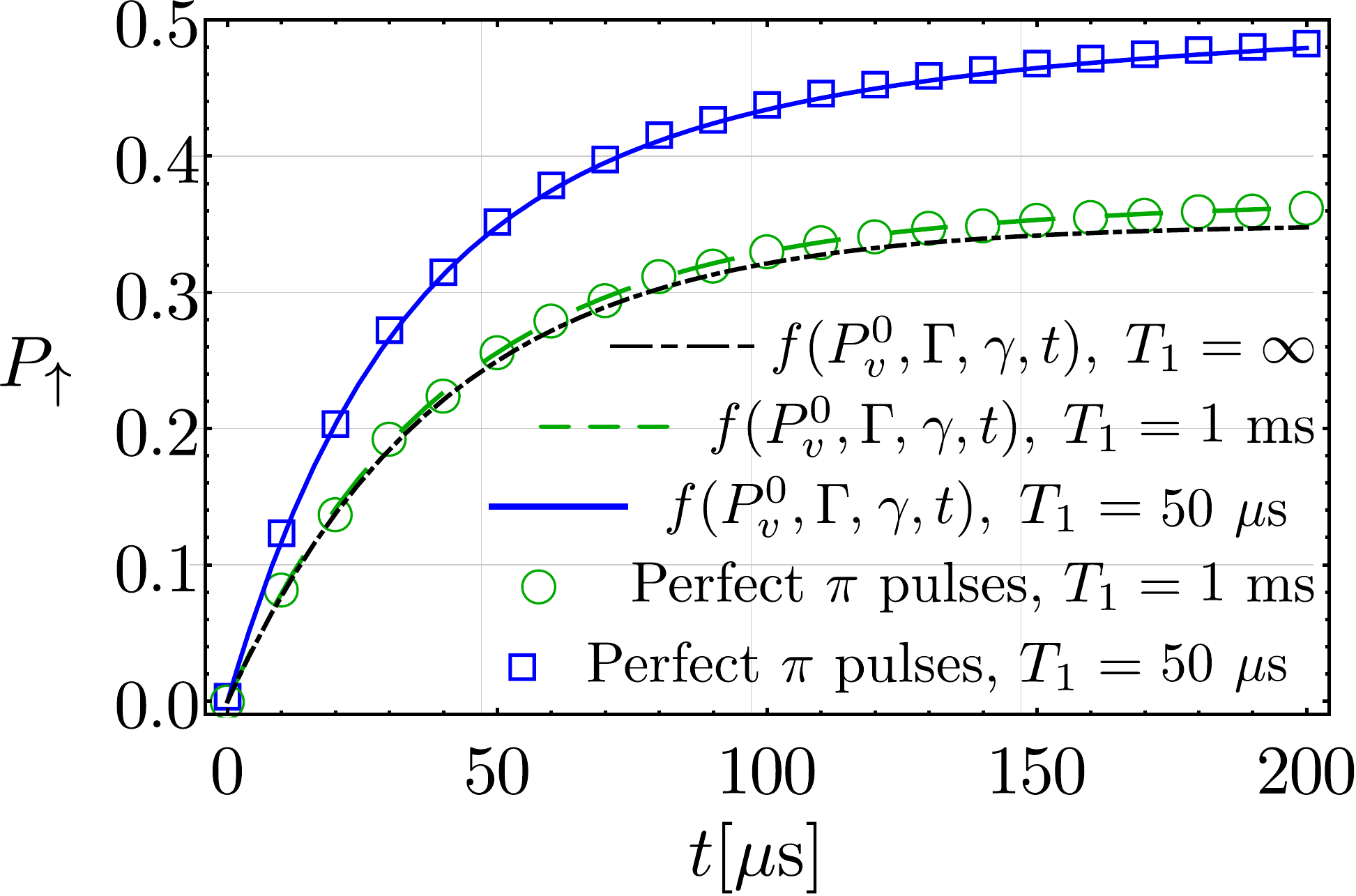}
\caption{(Color online) Spin-up probability after the echo sequence, when inter-valley scattering and spin relaxation are present. The parameters of the plot are the valley injection probabilities 
$P^0_v=0.7$ and $P^0_{\bar{v}}=0.3$, the inter-valley scattering rate $\Gamma=25\text{ kHz}$, the spin relaxation time  $T_1$ ($\gamma=1/T_1$), the external magnetic field $B_z=0.75\text{ T}$ and
the $z$-component of the magnetic field of the ferromagnet $B_z^{\rm FM}=-0.12\text { T}$.
The fitting function $f(P_{v}^0,\Gamma,\gamma,t)=0.5(1+P_v^0 e^{-(\Gamma+\gamma/2)t}+P_{\bar{v}}^0 e^{-\gamma t/2})$ was used.}
\label{Res3}
\end{figure}

\begin{figure}[t!]
\centering
\includegraphics[width=0.48\textwidth]{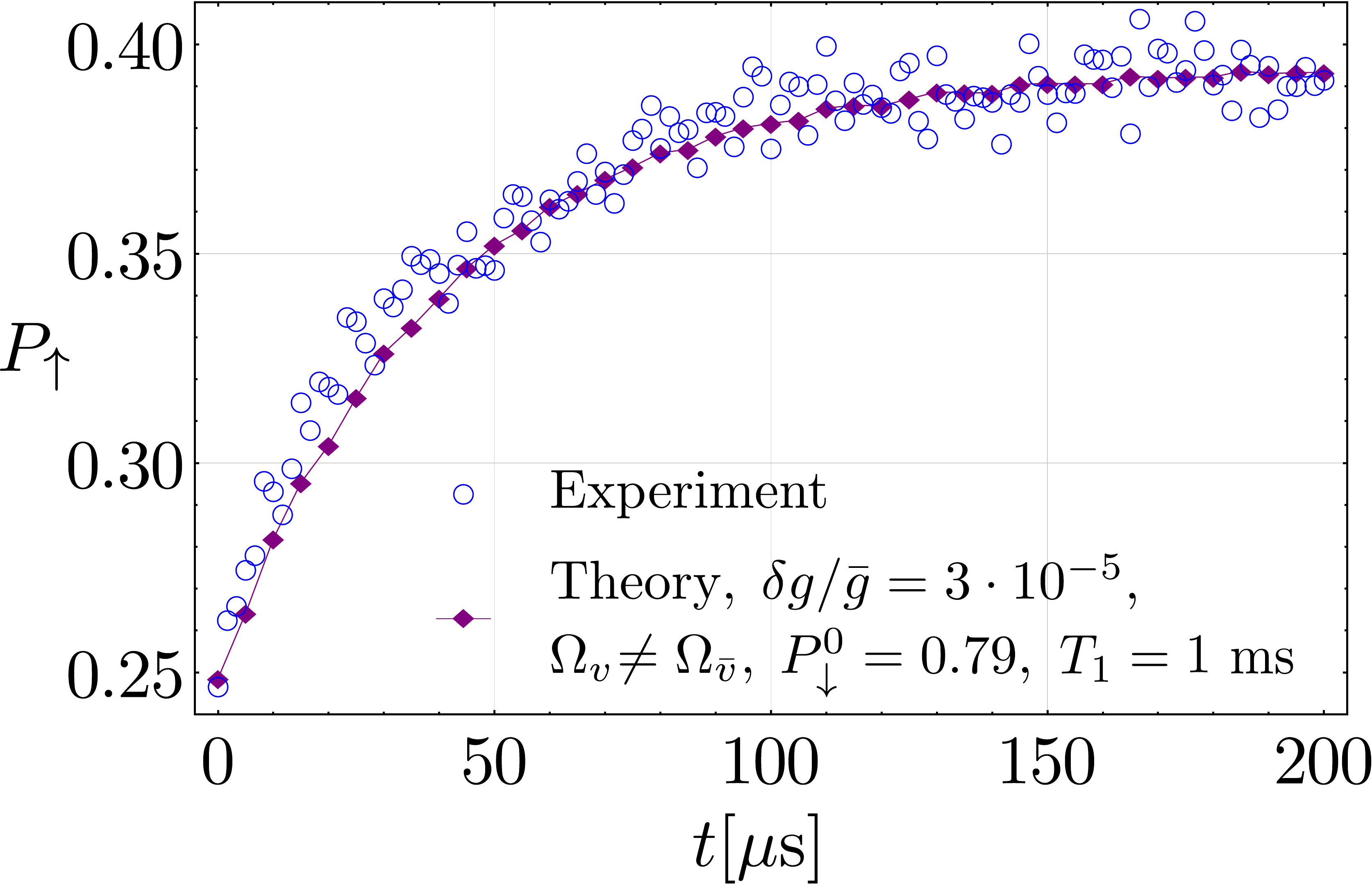}
\caption{(Color online) Probability $P_\uparrow$ for the echo sequence yielding the electron $|\!\uparrow\rangle$ state. The blue circles represent experimental findings [\onlinecite{Kawakami1}] and the purple diamonds are our theoretical findings
when the $\pi$ and $\pi/2$ pulses are imperfect and inter-valley and spin relaxations are present. 
The spin and valley injection probabilities are assumed to be $P_{v,\downarrow}^0=0.49$, $P_{v,\uparrow}^0=0.21$ and $P_{\bar{v},\downarrow}^0=0.3$.
The parameters of the plot are the external magnetic field $B_z=0.75\text{ T}$,
 the $z$-component of the magnetic field of the ferromagnet $B_z^{\rm FM}=-0.12\text{ T}$, valley dependent Rabi frequencies corresponding to the miscut angle $\theta\approx 0.3^{\circ}$, $\Omega_v=2\pi\cdot3.1\text{ MHz}$ and
 $\Omega_{\bar{v}}=2\pi\cdot3.7\text{ MHz}$, within the values suggested in a recent experimental study [\onlinecite{Kawakami1}].} 
\label{Res4}
\end{figure}

Because we are again interested in obtaining the echo envelope function as a measure of the coherence drop we will start from a Lindblad equation in a superoperator form
\begin{equation}\label{eq:Sup2}
\rho(t)=e^{\mathcal{L'}t}\rho(0).
\end{equation}
By repeating the procedure from Section \ref{mod} (Eq. (\ref{eq:SupEch}) and Eq. (\ref{eq:Proj})), 
we obtain the echo envelope function Fig. \ref{Res3} (probability that the electron spin is measured in the ${|\!\uparrow \rangle}$ state after a time $t$, when being subjected to a sequence of perfect $\pi/2-\pi-\pi/2$ pulses).
When the $g$-factor is valley dependent, the $\pi$ pulses perfect and electron spin relaxation is occurring the increase of the echo $P_\uparrow$ probability is caused by the interplay of valley and spin relaxations (Fig. \ref{Res3}, green circles and blue squares). 
The exponential function $f(P_{v}^0,\Gamma,\gamma,t)=0.5(1+P_v^0 e^{-(\Gamma+\gamma/2)t}+P_{\bar{v}}^0 e^{-\gamma t/2})$ describes the drop of coherence. In the $|v\rangle$ state the drop of coherence is caused by both spin and valley relaxation processes,  
while in the $|\bar{v}\rangle$ valley the drop of coherence is caused by spin relaxation processes. By comparing the results for $T_1=\infty$ (black dashed dotted line, Fig. \ref{Res3}) and $T_1=1\text{ ms}$ (green dashed line, Fig. \ref{Res3}), we see that the
spin relaxation happening on $T_1=1\text{ ms}$ timescales is increasing the $P_\uparrow$ probability by only $\sim 0.01$ on $\sim 200\text{ $\mu$s}$ timescales.

In Fig. \ref{Res4} we assume imperfect $\pi$ and $\pi/2$ pulses, with the rotation operators $R_x(\pi)=\exp{(-iH\pi/\hbar\Omega_v)}$ and $R_x(\pi/2)=\exp{(-iH\pi/\hbar2\Omega_v)}$, where the $H$ is given by Eq. (\ref{eq:Ham}), and $\Omega_{v}$ is the valley dependent 
Rabi frequency. During the free evolution stages the electron spin precesses around the external magnetic field. 
After the imperfect initial $\pi/2$ pulse, the electron spin is not perpendicular to the magnetic field, yielding rotations around the quantization axis
 with a frequency proportional to the Zeeman energy $g_{\bar{v}}\mu_{\rm B}(B_z+B_z^{\rm FM})/h$, where $B_z$ is the external magnetic field and $B_z^{\rm FM}$ is the $z$-component of the magnetic field of the ferromagnet. 
For $B_z=0.75\text{ T}$ and $B_z^{\rm FM}=-0.12\text{ T}$ this oscillations happen on $\sim50\text{ ps}$ timescale, 
with the amplitude of the oscillations being given by the valley dependent Rabi frequencies $\Omega_v$ and $\Omega_{\bar{v}}$ and Rabi dependent $g$-factors $g_v$ and $g_{\bar{v}}$. 
Therefore, the $P_\uparrow$ probability is very sensitive to the duration of the free evolution stage. The relaxation time $T_1=1\text{ ms}$ is within the value suggested in a recent experimental study. 

By comparing experimental data points (blue circles) with the result of our modeling (purple diamonds) we conclude that the saturation value of the $P_\uparrow$ probability 
$P_\uparrow(t\rightarrow\infty)\approx 0.39$ and the $P_\uparrow$ probability close to $t=0$, $P_\uparrow(t=0)\approx 0.25$ are all within the the values measured in a recent experimental study [\onlinecite{Kawakami1}] and that our model yields the correct 
functional form of $P_\uparrow$ probability increase.

\section{CONCLUSION}\label{Con}

To conclude, we have discussed the control of the electron spin inside a Si/SiGe quantum dot with a ferromagnet embedded on top. The stray magnetic field of the ferromagnet combined with Si/SiGe interference imperfections 
consequently leads to a valley dependent $g$-factor. 
When a valley dependent $g$-factor, alongside with valley relaxation is present, a novel decoherence mechanism exists, further limiting the coherence of the electron spin. Furthermore, the control of the electron spin state on the Bloch sphere is influenced
 by a valley dependent $g$-factor and Rabi frequency. Our model gives a good qualitative and quantitative description of recent experimental studies. 
Further research on this topic will move towards including the drop of coherence due to the presence of nuclear spins.

\section{Acknowledgments}\label{Acknowledgment}

We thank Erika Kawakami and Pasquale Scarlino for giving us access to their experimental data and providing additional information about their experiment. 
Furthermore, we thank Heng Wang and Niklas Rohling for useful discussions and the European Union within the S$^3$NANO Marie Curie ITN and the DFG within the SFB767 for financial support.
\bibliographystyle{apsrev}
\bibliography{SiGe_EDSR}

\end{document}